# Tunnelling Through Time Series: A Probabilistic Visibility Graph for Local and Global Pattern Discovery

ROBERTO C. SOTERO*

Department of Radiology, and Hotchkiss Brain Institute, University of Calgary, AB, Canada

JOSE M. SANCHEZ-BORNOT

Intelligent Systems Research Centre, Ulster University, Derry Londonderry, United Kingdom

The growing availability of high-resolution, long-term time series data has highlighted the need for methods capable of capturing both local and global patterns. To address this, we introduce the Probabilistic Visibility Graph (PVG), a novel approach inspired by the quantum tunnelling phenomenon. The PVG extends the classical Visibility Graph (VG) by introducing probabilistic connections between time points that are obstructed in the VG due to intermediate values. We demonstrate the PVG's effectiveness in capturing long-range dependencies through simulations of amplitude-modulated signals and analysis of electrocorticography (ECoG) data under rest and anesthesia conditions. Key results show that the PVG presents distinct network properties between rest and anesthesia, with rest exhibiting stronger small-worldness and scale-free behavior, reflecting a hub-dominated, centralized connectivity structure, compared to anesthesia. These findings highlight the PVG's potential for analyzing complex signals with interacting temporal scales, offering new insights into neural dynamics and other real-world phenomena.

## 1 INTRODUCTION

Modern science increasingly relies on high-resolution, long-duration signals to reveal intricate local and global temporal dynamics essential for understanding complex phenomena such as brain activity [1], market trends [2], or ecological changes [3]. However, traditional time series analysis methods, which primarily emphasize local structures and short-term interactions [4], are frequently inadequate for handling such data. This has driven the development of advanced techniques capable of capturing both local and global patterns, as well as long-range dependencies, in time series [5].

Over the past two decades, complex network approaches have emerged as powerful tools for analyzing nonlinear time series, offering insights into the underlying structure and dynamics of complex systems [6]. By transforming time series data into networks, these methods enable the application of graph-theoretical measures to uncover patterns, relationships, and anomalies that are often obscured in traditional time-domain or frequency-domain analyses. Various network-based approaches have been developed, including recurrence networks [7], transition networks [8], and the Visibility Graph (VG) [9], each tailored to specific characteristics of time series data. Among these approaches, the VG has gained significant attention due to its simplicity, interpretability, and ability to map time series into networks while preserving their intrinsic properties. Introduced by Lacasa et al. [9], the VG method constructs a graph where each data point in the time series is represented as a node, and edges are drawn between nodes if the corresponding data points are "visible" to each other—that is, if a straight line connecting them does not intersect any intermediate data points. This geometric interpretation allows the VG to capture important features of the time series, making it particularly effective for analyzing nonlinear and non-stationary data [6].

* Correspondence: roberto.soterodiaz@ucalgary.ca

Despite its success, the classical VG [9] has limitations, particularly in capturing long-range dependencies in time series. The VG primarily emphasizes the local structure of the time series, as visibility between two points is determined by the heights of intermediate points. As the time window between points increases, the likelihood of obstructions between them also rises. This local focus restricts the VG's ability to capture long-range dependencies, which are essential for understanding the dynamics of many real-world systems. For instance, brain activity often comprises slow-frequency rhythms that are critical for understanding cognitive processes [10], [11].

To address these limitations, we introduce the Probabilistic Visibility Graph (PVG), a novel approach that extends the classical VG by incorporating probabilistic connections inspired by the quantum tunneling phenomenon. Here we explore the theoretical foundations and practical applications of the PVG method, highlighting its advantages over the classical VG when dealing with time series data with complex local and global patterns.

## 2  METHODS

The classical VG [9] transforms a time series into a graph where each data point $x(t_i)$ is represented as a node, and edges are drawn between nodes $i$ and $j$ if the corresponding data points are "visible" to each other. Visibility is determined by the geometric criterion:

$$x(t_n) < x(t_i) + \frac{x(t_j)-x(t_i)}{t_j-t_i}(t_n - t_i), \forall t_n \in (t_i, t_j) \tag{1}$$

This criterion ensures that a straight line connecting $x(t_i)$ and $x(t_j)$ does not intersect any intermediate data points $x(t_n)$. The adjacency matrix $A$ of the VG is defined as:

$$A_{ij} = \begin{cases} 1 & \text{if } x(t_i) \text{ and } x(t_j) \text{ are visible} \\ 0 & \text{otherwise} \end{cases} \tag{2}$$

In quantum mechanics, particles can "tunnel" through potential barriers with a certain probability, even when their energy is insufficient to overcome the barrier classically [12]. Drawing inspiration from this phenomenon, the PVG method introduces probabilistic connections between nodes in the visibility graph, even when the corresponding time series points are obstructed by intermediate points. This allows the PVG to effectively "tunnel" through the time series, capturing connections that would be absent in a classical VG. By incorporating this probabilistic framework, the PVG is capable of modeling both local and global patterns making it well suited for analyzing complex systems.

The PVG is constructed as follows. First, to ensure consistency, the input time series $x(t)$ is normalized to the range [0,1] as:

$$\tilde{x}(t) = \frac{x(t)-min(x)}{max(x)-min(x)} \tag{3}$$

where $max(x)$ and $min(x)$ are the maximum and minimum values of the time series, respectively. Then, we introduce **probabilistic connections** $P_{ij}$ between nodes based on the height of intermediate obstructions:

$$P_{ij} = e^{-\rho h_{max}} \tag{4}$$

where the decay parameter $\rho$ controls the influence of obstructions. Large $\rho$ values penalize non-visible connections more heavily, allowing the tuning of the PVG's sensitivity to long-range connections. $h_{max}$ is the maximum obstruction height between nodes *i* and *j*.

The obstruction height $h_n$ at an intermediate point *n* is defined as:

$$h_n = \tilde{x}(t_n) - \left(\tilde{x}(t_i) + \frac{\tilde{x}(t_j) - \tilde{x}(t_i)}{t_j - t_i}(t_n - t_i)\right) \tag{5}$$

Between nodes *i* and *j*, there can be multiple obstructions, each contributing an obstruction height $h_n$. The maximum obstruction height $h_{max}$ is defined as the largest obstruction height among all intermediate points. If no obstruction exists $h_{max} = 0$, resulting in a connection probability $P_{ij} = 1$.

In addition to the connection probability $P_{ij}$, the PVG incorporates an interaction strength $W_{ij}$ between nodes *i* and *j* ($t_j > t_i$) that has been previously introduced for the VG [13]:

$$W_{ij} = arctan\left(\frac{|x(t_j) - x(t_i)|}{t_j - t_i}\right) \tag{6}$$

Here $W_{ij}$ quantifies the strength of the interaction between nodes based on both the value difference and the temporal distance between the corresponding time points. The weighted connectivity matrix $M$ of the PVG is then defined as the elementwise product of the strength matrix $W$ and the probability matrix $P$:

$$M_{ij} = W_{ij} \cdot P_{ij} \tag{7}$$

The connectivity $M_{ij}$ will be zero in two cases: (1) when $W_{ij} = 0$, which occurs if $x(t_i) = x(t_j)$, and (2) when $P_{ij} = 0$ which only happens if $h_{max} = \infty$. Note that this formulation ensures that all points in the time series have a non-zero probability of being connected, even if the probability is very small. While in some cases it may be desirable to connect all points in the time series, in most practical scenarios, it is more meaningful to introduce a threshold probability $P_0$. Connections with $P_{ij} \geq P_0$ are included in the graph, while those with $P_{ij} < P_0$ are excluded. The adjacency matrix $A_{PVG}$ of the PVG is then defined as:

$$A_{PVG,ij} = \begin{cases} 1 & \text{if } P_{ij} \geq P_0 \\ 0 & \text{otherwise} \end{cases} \tag{8}$$

When $P_0 = 0$, the PVG includes interactions between all the time points. Conversely, when $P_0 = 1$, the PVG reduces to the classical VG. Like the classical VG, the construction of the PVG is computationally intensive since it requires checking for the obstructions between all pairs of nodes. This results in a computational cost of $O(N^2)$.

## 2.1 Construction of the PVG

Figure 1 illustrates the construction and analysis of the PVG applied to a normalized time series. In subplot (a), the normalized time series is shown, with two specific time points (*i* and *j*) highlighted and connected through a dashed green line to demonstrate the probabilistic connection $P_{ij}$ between them. Obstructions between the time points are visualized in magenta with the maximum obstruction height $h_{max}$ highlighted in black. Figure. 1(b) shows the graph representation of the PVG, contrasting connections absent in the classical VG (green edges) with those present in both VG and PVG (blue edges). The specific connection between nodes *i* and *j* analyzed in Figure. 1(a) is highlighted in darker green, emphasizing its probabilistic nature. Figure. 1(c) shows the weighted connectivity matrix *C*. The matrix presents the strength of connections between nodes, with white regions indicating no connection after a threshold of $P_0 = 0.5$ was used, and colored regions representing weighted connectivity. Together, Figure. 1 subplots demonstrate the PVG's ability to capture long-range connections in the time series and provide a quantitative view of the associated network's connectivity.

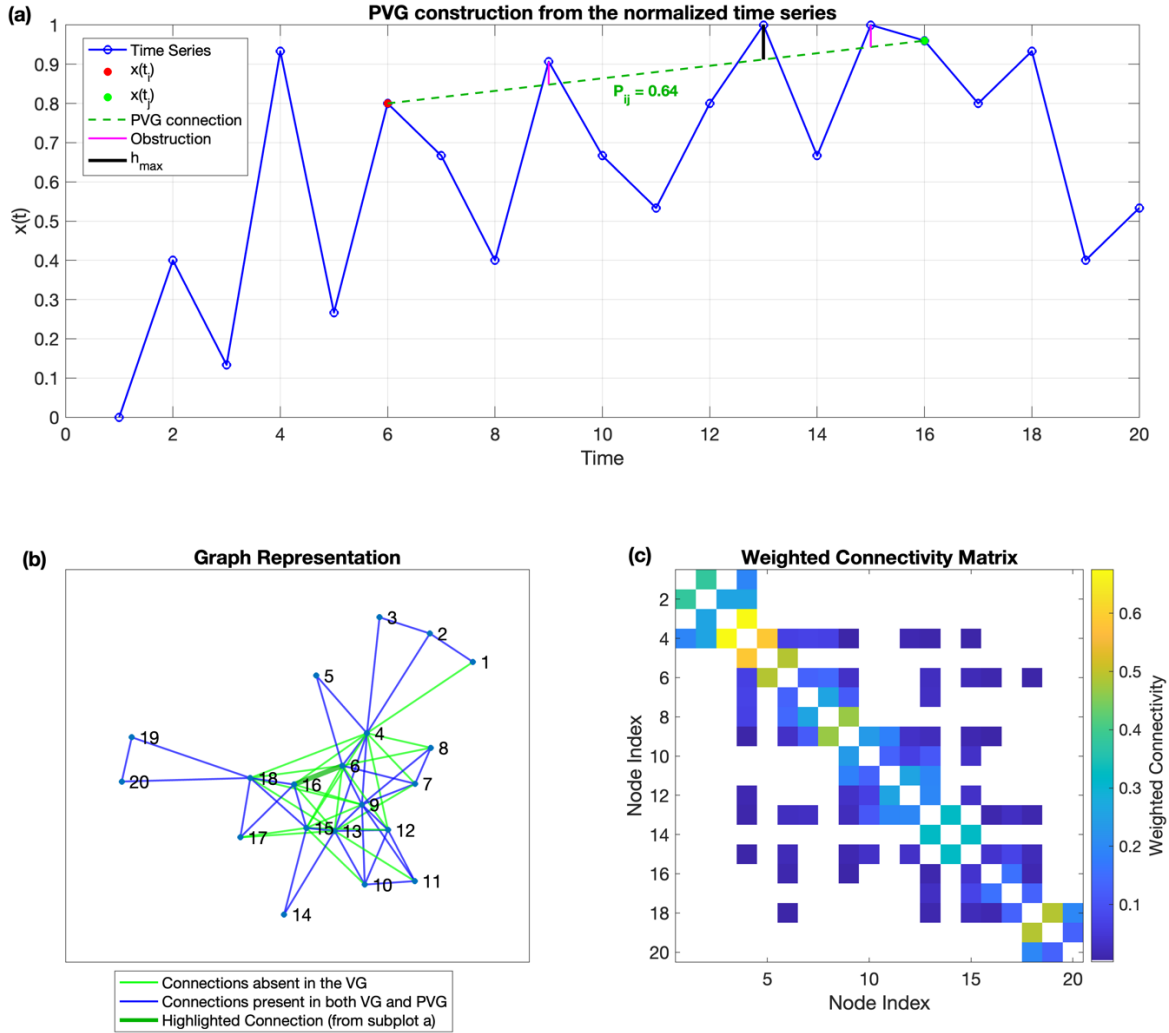

Figure 1: Probabilistic Visibility Graph (PVG) construction. (a) Normalized time series. Nodes $i$ and $j$ (red and green markers, respectively) are connected with probability $P_{ij}$ (green dashed line). Obstructions between nodes are shown in magenta, with the maximum obstruction height $h_{max}$ highlighted in black. (b) Graph representation of the PVG. (c) Weighted connectivity matrix. White regions indicate no connection, while colored regions represent the strength of weighted connections.

## 2.2 Analysis of a simulated Amplitude-Modulated (AM) signal

To further explore how the PVG works, we generated an AM signal $x(t)$ as:

$$x(t) = A_c\big(1 + m\cos(2\pi f_m t)\big)\sin(2\pi f_c t) + \eta(t) \quad (9)$$

where $A_c$ = 1 is the amplitude of the carrier wave, $f_c$ = 40 $Hz$ is the frequency of the fast oscillations (carrier frequency), $f_m$ = 6 $Hz$ is the frequency of the slow modulation, $m$ = 0.5 is the modulation depth, and $\eta(t)$ = 0.01 is additive Gaussian noise. This signal allows us to study how the PVG captures interactions between local (high-frequency) and global (low-frequency) patterns, which are common in many real-world systems. Notably, the phenomenon of phase-amplitude coupling (PAC) in neurophysiological signals, which has attracted significant attention in recent years, can be modeled as an AM signal. PAC refers to the interaction between the phase of a low-frequency rhythm (such as a 6 Hz theta oscillation) and the amplitude of a high-frequency rhythm (such as a 40 Hz gamma oscillation), and it has been widely studied in applications such as understanding brain dynamics [14], [15], epileptic seizure localization [16], and characterizing cognitive processes [17]. By using an AM signal, we mimic this biologically relevant phenomenon, providing a meaningful test case for the PVG.

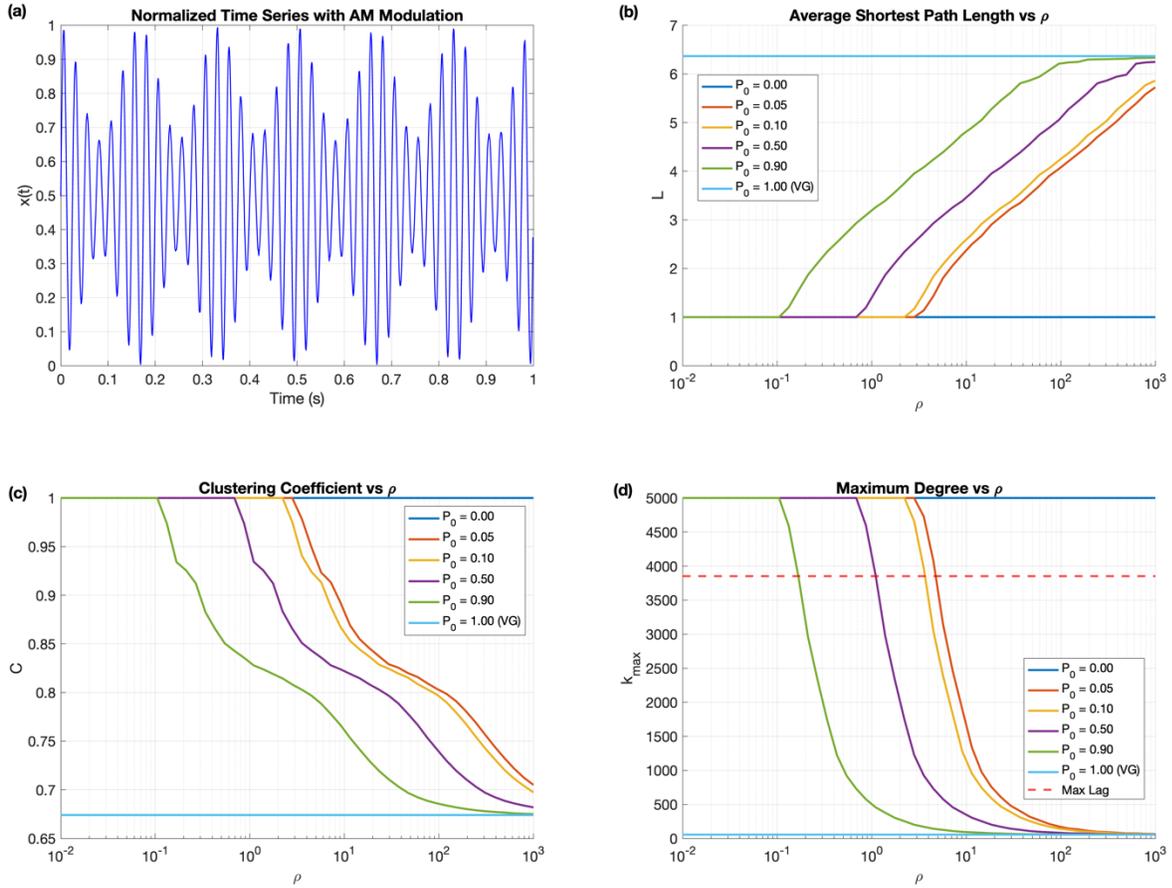

Figure 2: Analysis of the PVG for an amplitude-modulated (AM) time series. (a) Normalized AM time series generated with a carrier frequency $f_c = 40\ Hz$, modulation frequency $f_m = 6\ Hz$, and modulation depth $m = 0.5$. The time series includes additive noise with an amplitude of 0.01. (b) Average shortest path length $L$ (b), clustering coefficient $C$ (c), and maximum degree $k_{max}$ (d) of the PVG, as a function of the decay parameter $\rho$ (log scale).

Figure 2(a) displays a 1-second window of the normalized 5-second AM signal, constructed with a temporal resolution of 0.001 seconds. To analyze the topological properties of the associated PVG, we computed three key network measures using the Brain Connectivity Toolbox [18]: the average shortest path length ($L$), reflecting the global efficiency of information transfer; the average clustering coefficient ($C$), quantifying the local connectivity or "cliquishness" of the network; and the maximum degree ($k_{max}$), representing the highest degree among all nodes and identifying the connectivity of the most central node (hub). These measures were computed from the binary adjacency matrix $A_{PVG}$ to avoid complications from weighted connections, which can obscure the interpretation of the results.

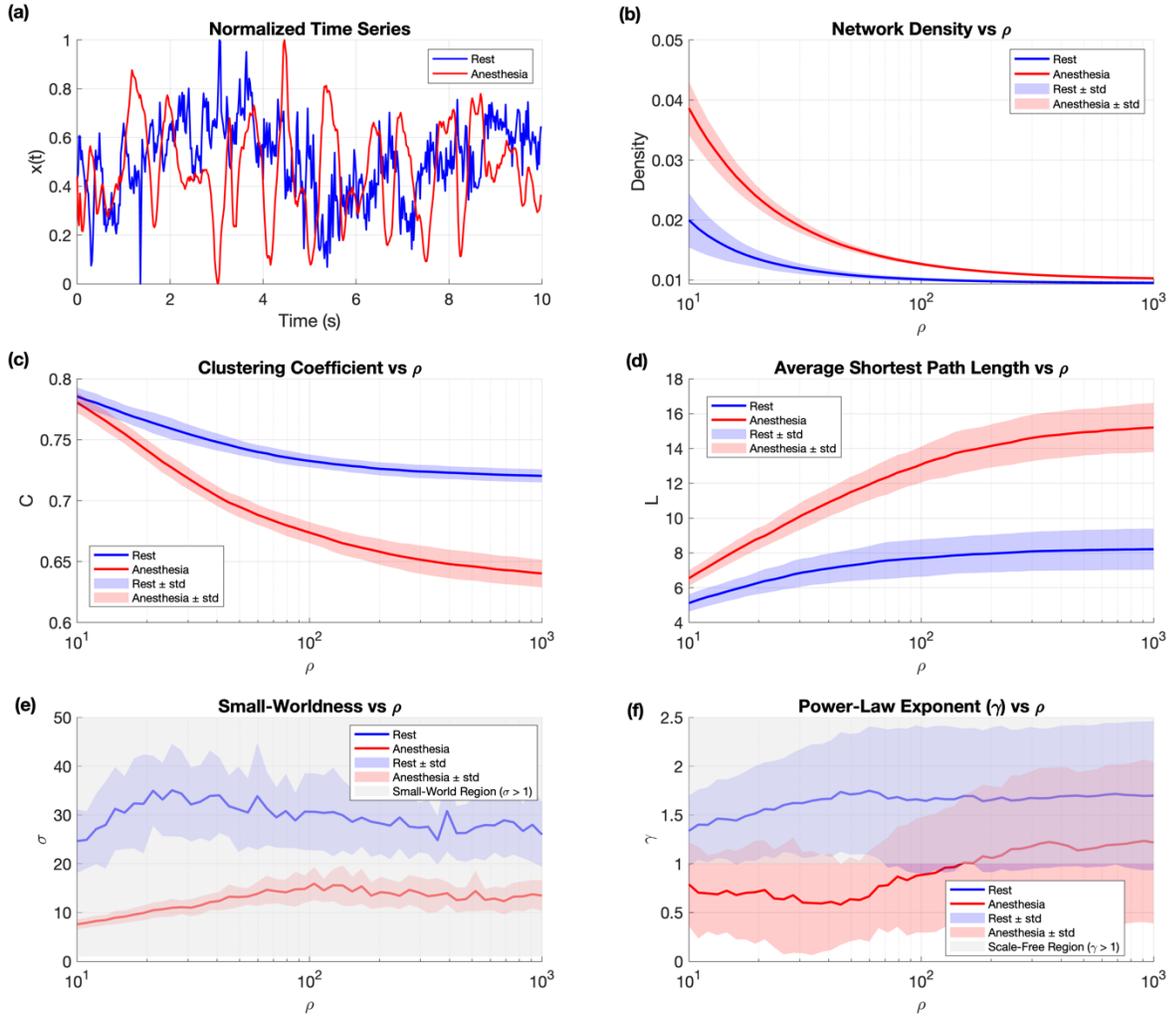

**Figure 3:** Probabilistic Visibility Graph (PVG) analysis of ECoG data under rest (blue) and anesthesia (red) conditions. (a) Normalized time series from a premotor cortex electrode. Average degree (b), clustering coefficient (c), average path length (d), Small-worldness parameter (e), and power-law exponent of the degree distribution (f) of the PVG as a function of $\rho$ (log scale). All metrics represent means across 30 segments, with shaded areas indicating standard deviation.

Figure 2(b), (c), and (d), display $L$, $C$, and $k_{max}$ as a function of the decay parameter $\rho$ for different threshold probabilities $P_0$, respectively. Figure 2(b) shows that as $\rho$ increases, $L$ also increases, reflecting the PVG's probabilistic framework: larger $\rho$ values reduce the probability of direct connections between distant nodes, resulting in longer paths. The classical VG ($P_0 = 1$) exhibits the largest $L$, as it only includes deterministic connections without probabilistic tunneling. In contrast, lower $P_0$ values result in shorter paths due to the inclusion of more probabilistic connections. Figure 2(c) shows that higher $\rho$ values reduce $C$, as they decrease the likelihood of forming local connections. Conversely, lower $P_0$ values enhance local "cliquishness", increasing $C$. Finally, in Figure 2(d) we found that as $\rho$ increases, $k_{max}$ decreases. The classical VG exhibits the smallest value ($k_{max} = 56$), significantly lower than the values achievable by the PVG across different $\rho$ values. Specifically, the PVG's $k_{max}$ ranges from the theoretical maximum of 5000 (for small $\rho$) down to the classical VG's value of 56 for very large $\rho$. For comparison, the maximum lag derived from the autocorrelation function of the time series was 3852, underscoring the classical VG's inability to capture long-range interactions. In contrast, the PVG effectively captures these interactions across a wide range of $\rho$ values. Additionally, lower $P_0$ values resulted in the expected increase in $k_{max}$. In summary, Figure 2 highlights the PVG's superior ability to capture long-range interactions compared to the classical VG, while also demonstrating its flexibility in balancing local and global connectivity patterns through variations in $\rho$ and $P_0$.

### 2.3 Analysis of real electrocorticography (ECoG) data

To evaluate the PVG with real-world data, we analyzed ECoG recordings from a macaque monkey during both rest and anesthesia conditions [19]. The ECoG array comprised 128 electrodes with signals sampled at a rate of 1000 Hz. To avoid aliasing during downsampling, we first applied a low-pass Butterworth filter with a cutoff frequency of 25 Hz (the Nyquist frequency for the target sampling rate of 50 Hz). The time series were then divided into 30 non-overlapping segments, each lasting 10 seconds, and downsampled to a 50 Hz sampling rate to streamline subsequent analysis [20].

Figure 3(a) displays the normalized time series for both rest (blue) and anesthesia (red) conditions for one segment. The time series under anesthesia shows reduced variability compared to the rest condition, reflecting the suppression of neural activity during anesthesia. Figures 3(b), (c), and (d) present the network density, clustering coefficient $C$, and average shortest path length $L$ as functions of $\rho$, respectively, after averaging across the 30 segments. In Figure 3(b), as $\rho$ increases, the network density decreases with the anesthesia condition exhibiting higher network density than the rest condition. Figure 3(c) reveals that higher $\rho$ values reduce $C$, with the rest condition having higher $C$ compared to anesthesia. On the other hand, Figure 3(d) shows that higher $\rho$ values lead to longer paths with the anesthesia condition exhibits longer paths compared to rest. Fig. 3(e) explores the small-worldness ($\sigma$) of the PVG, computed as [21]:

$$\sigma = \frac{C/C_{rand}}{L/L_{rand}} \qquad (10)$$

where $C_{rand}$ and $L_{rand}$ are the clustering coefficient and average shortest path length, respectively, for a random network with the same number of nodes and edges as $A_{PVG}$. A network is considered small-world if $\sigma > 1$, which occur when it has similar $L$ but significantly higher $C$ compared to its random counterpart [21]. Both rest and anesthesia conditions demonstrated small-world properties with rest showing significantly higher $\sigma$ than the rest condition.

Figure 3(f) presents the power-law exponent $\gamma$ of the degree distribution $P(k)$ as a function of $k$. In scale-free networks, $P(k)$ follows a power-law $P(k) \sim k^{-\gamma}$ [22], with networks considered scale-free when $\gamma > 1$ [23], with a more restrictive definition requiring $2 < \gamma < 3$ [24]. Here $\gamma$ was estimated by fitting a linear regression to the log-transformed degree distribution [25]. The rest condition exhibited scale-free behavior across all explored $\rho$ values, indicating the presence of a hub-dominated network structure. In contrast, the anesthesia condition showed a less clear scale-free behavior that started around $\rho = 200$. Overall, our findings suggest that anesthesia induced a reorganization of the PVG, characterized by a more homogeneous and less structured state with fewer highly connected nodes. This reorganization reflects the disruption of normal brain dynamics during anesthesia.

### 3 CONCLUSIONS

In this paper, we introduced the PVG, a novel framework inspired by the quantum tunnelling phenomenon, for analyzing time series data. The PVG extends the VG by incorporating probabilistic connections, enabling the capture of both local and global patterns in complex temporal data. Through theoretical analysis and application to real-world ECoG data under rest and anesthesia conditions, we demonstrated the PVG's ability to model long-range interactions and reveal condition-specific network properties. Key findings include the PVG's sensitivity to the decay parameter $\rho$ and threshold probability $P_0$, its ability to exhibit small-world and scale-free properties under specific conditions, and its superior performance over the classical VG in capturing long-range dependencies. These results highlight the PVG's potential as a powerful tool for analyzing complex time series, offering new insights into the topological reorganization of their associated networks under different physiological states.

## ACKNOWLEDGMENTS

This work was supported by grant RGPIN-2022-03042 from Natural Sciences and Engineering Council of Canada. The authors are grateful for access to the Tier 2 High-Performance Computing resources provided by the Northern Ireland High Performance Computing (NI-HPC) facility funded by the Engineering and Physical Sciences Research Council (EPSRC), Grant No. EP/T022175/1.

## REFERENCES

[1] T. D. Lagerlund, G. D. Cascino, K. M. Cicora, and F. W. Sharbrough, "Long-Term Electroencephalographic Monitoring for Diagnosis and Management of Seizures," *Mayo Clinic Proceedings*, vol. 71, no. 10, pp. 1000–1006, Oct. 1996, doi: 10.4065/71.10.1000.
[2] Z. Dong, X. Fan, and Z. Peng, "FNSPID: A Comprehensive Financial News Dataset in Time Series," in *Proceedings of the 30th ACM SIGKDD Conference on Knowledge Discovery and Data Mining*, in KDD '24. New York, NY, USA: Association for Computing Machinery, Aug. 2024, pp. 4918–4927. doi: 10.1145/3637528.3671629.
[3] C. Vancutsem *et al.*, "Long-term (1990–2019) monitoring of forest cover changes in the humid tropics," *Science Advances*, vol. 7, no. 10, p. eabe1603, Mar. 2021, doi: 10.1126/sciadv.abe1603.
[4] "Analysis of Financial Time Series, 3rd Edition | Wiley," Wiley.com. Accessed: Jan. 21, 2025. [Online]. Available: https://www.wiley.com/en-us/Analysis+of+Financial+Time+Series%2C+3rd+Edition-p-9780470414354
[5] Y. Liu, C. Gong, L. Yang, and Y. Chen, "DSTP-RNN: A dual-stage two-phase attention-based recurrent neural network for long-term and multivariate time series prediction," *Expert Systems with Applications*, vol. 143, p. 113082, Apr. 2020, doi: 10.1016/j.eswa.2019.113082.
[6] Y. Zou, R. V. Donner, N. Marwan, J. F. Donges, and J. Kurths, "Complex network approaches to nonlinear time series analysis," *Physics Reports*, vol. 787, pp. 1–97, Jan. 2019, doi: 10.1016/j.physrep.2018.10.005.
[7] J.-P. Eckmann, S. O. Kamphorst, and D. Ruelle, "Recurrence Plots of Dynamical Systems," in *Turbulence, Strange Attractors and Chaos*, vol. Volume 16, in World Scientific Series on Nonlinear Science Series A, no. Volume 16, vol. Volume 16. , WORLD SCIENTIFIC, 1995, pp. 441–445. doi: 10.1142/9789812833709_0030.
[8] H. Guo, J.-Y. Zhang, Y. Zou, and S.-G. Guan, "Cross and joint ordinal partition transition networks for multivariate time series analysis," *Front. Phys.*, vol. 13, no. 5, pp. 1–10, Oct. 2018, doi: 10.1007/s11467-018-0805-0.
[9] L. Lacasa, B. Luque, F. Ballesteros, J. Luque, and J. C. Nuño, "From time series to complex networks: The visibility graph," *Proceedings of the National Academy of Sciences*, vol. 105, no. 13, pp. 4972–4975, Apr. 2008, doi: 10.1073/pnas.0709247105.
[10] E. Tan *et al.*, "Theta activity and cognitive functioning: Integrating evidence from resting-state and task-related developmental electroencephalography (EEG) research," *Dev Cogn Neurosci*, vol. 67, p. 101404, Jun. 2024, doi: 10.1016/j.dcn.2024.101404.
[11] T. Harmony, "The functional significance of delta oscillations in cognitive processing," *Front. Integr. Neurosci.*, vol. 7, Dec. 2013, doi: 10.3389/fnint.2013.00083.
[12] D. J. Griffiths and D. F. Schroeter, "Introduction to Quantum Mechanics," Higher Education from Cambridge University Press. Accessed: Jan. 28, 2025. [Online]. Available: https://www.cambridge.org/highereducation/books/introduction-to-quantum-mechanics/990799CA07A83FC5312402AF6860311E
[13] S. Supriya, S. Siuly, H. Wang, J. Cao, and Y. Zhang, "Weighted Visibility Graph With Complex Network Features in the Detection of Epilepsy," *IEEE Access*, vol. 4, pp. 6554–6566, 2016, doi: 10.1109/ACCESS.2016.2612242.
[14] R. C. Sotero *et al.*, "Laminar distribution of phase-amplitude coupling of spontaneous current sources and sinks," *Frontiers in neuroscience*, vol. 9, p. 454, 2015.
[15] R. C. Sotero, "Topology, cross-frequency, and same-frequency band interactions shape the generation of phase-amplitude coupling in a neural mass model of a cortical column," *PLoS computational biology*, vol. 12, no. 11, p. e1005180, 2016.
[16] N. E. Cámpora, C. J. Mininni, S. Kochen, and S. E. Lew, "Seizure localization using pre ictal phase-amplitude coupling in intracranial electroencephalography," *Sci Rep*, vol. 9, no. 1, p. 20022, Dec. 2019, doi: 10.1038/s41598-019-56548-y.
[17] J. Daume, T. Gruber, A. K. Engel, and U. Friese, "Phase-Amplitude Coupling and Long-Range Phase Synchronization Reveal Frontotemporal Interactions during Visual Working Memory," *J. Neurosci.*, vol. 37, no. 2, pp. 313–322, Jan. 2017, doi: 10.1523/JNEUROSCI.2130-16.2016.
[18] M. Rubinov and O. Sporns, "Complex network measures of brain connectivity: Uses and interpretations," *NeuroImage*, vol. 52, no. 3, pp. 1059–1069, Sep. 2010, doi: 10.1016/j.neuroimage.2009.10.003.
[19] N. Oosugi, K. Kitajo, N. Hasegawa, Y. Nagasaka, K. Okanoya, and N. Fujii, "A new method for quantifying the performance of EEG blind source separation algorithms by referencing a simultaneously recorded ECoG signal," *Neural Networks*, vol. 93, pp. 1–6, Sep. 2017, doi: 10.1016/j.neunet.2017.01.005.
[20] R. C. Sotero and J. Sanchez-Bornot, "Estimating the Excitatory-Inhibitory Balance from Electrocorticography Data using Physics-Informed Neural Networks," in *Proceedings of the 11th International Conference on Bioinformatics Research and Applications*, in ICBRA '24. New York, NY, USA: Association for Computing Machinery, Jan. 2025, pp. 113–118. doi: 10.1145/3700666.3700701.
[21] M. D. Humphries and K. Gurney, "Network 'Small-World-Ness': A Quantitative Method for Determining Canonical Network Equivalence," *PLOS ONE*, vol. 3, no. 4, p. e0002051, Apr. 2008, doi: 10.1371/journal.pone.0002051.

[22] R. Albert and A.-L. Barabási, "Statistical mechanics of complex networks," *Rev. Mod. Phys.*, vol. 74, no. 1, pp. 47–97, Jan. 2002, doi: 10.1103/RevModPhys.74.47.
[23] A. D. Broido and A. Clauset, "Scale-free networks are rare," *Nat Commun*, vol. 10, no. 1, p. 1017, Mar. 2019, doi: 10.1038/s41467-019-08746-5.
[24] S. N. Dorogovtsev and J. F. F. Mendes, "Evolution of networks," *Advances in Physics*, vol. 51, no. 4, pp. 1079–1187, Jun. 2002, doi: 10.1080/00018730110112519.
[25] M. Newman, "Power laws, Pareto distributions and Zipf's law," *Contemporary Physics*, vol. 46, no. 5, pp. 323–351, Sep. 2005, doi: 10.1080/00107510500052444.